\documentclass[aps,prl,twocolumn,superscriptaddress,showpacs]{revtex4}

\usepackage[dvips]{graphicx}
\usepackage{array}

\usepackage{amsmath}
\usepackage{amssymb}

\def\bdm{\begin{displaymath}} \def\edm{\end{displaymath}}
\def\nn{\nonumber} \def\bc{\begin{center}} \def\ec{\end{center}}
\def\be{\begin{equation}} \def\ee{\end{equation}}

 \def\vone{{\bf 1}}   
   
\def\vd{{\bf d}}   \def\ve{{\bf e}}
   
\def\vg{{\bf g}}   
   
\def\vk{{\bf k}}   
   
   \def\vone{{\bf 1}}

  \def\v0{{\bf 0}} 
  
  \def\NF{N_{\rm F}} 
   
\def\vgamma{{\mbox{\boldmath$\gamma$}}} 
\def\vDelta{{\mbox{\boldmath$\Delta$}}} \def\vtau{{\mbox{\boldmath$\tau$}}}

\begin{document}


\title{Electronic Raman scattering in non-centrosymmetric superconductors}



\author{Ludwig Klam}
\email[]{L.Klam@fkf.mpg.de}
\affiliation{Max-Planck-Institut f\"{u}r Festk\"{o}rperforschung, Heisenbergstrasse 1, D-70569 Stuttgart}

\author{Dietrich Einzel}
\affiliation{Walther-Meissner-Institut, Bayerische Akademie der Wissenschaften, D-85748 Garching}

\author{Dirk Manske}
\affiliation{Max-Planck-Institut f\"{u}r Festk\"{o}rperforschung, Heisenbergstrasse 1, D-70569 Stuttgart}


\date{\today}

\begin{abstract} 
We formulate a theory for the polarization--dependence of the electronic (pair--breaking) Raman response for the recently discovered non--centrosymmetric superconductors in the clean limit at zero temperature. Possible applications include the systems CePt$_3$Si and Li$_2$Pd$_x$Pt$_{3-x}$B which reflect the two important classes of the involved spin--orbit coupling.
We provide analytical expressions for the Raman vertices for these two classes and calculate the polarization
dependence of the electronic spectra. We predict a two--peak structure and different power laws with respect to the unknown relative magnitude of the singlet and triplet contributions to the superconducting order parameter, revealing a large variety
of characteristic fingerprints of the underlying condensate.
\end{abstract}


\maketitle

The order parameter of conventional and unconventional superconductors is usually classified as either spin singlet (even parity) or spin triplet (odd parity) by the Pauli exclusion principle. A necessary prerequisite for such a classification is, however,
the existence of an inversion center. Something of a stir has been caused by the discovery of the bulk
superconductor CePt$_3$Si without inversion symmetry \cite{Bauer:2004:01}, which initiated extensive theoretical \cite{Frigeri:2004:02,Samokhin:2008:01} and experimental \cite{Bauer:2005:02,Fak:2008:01} studies. In such systems the existence of an antisymmetric potential gradient causes a parity--breaking antisymmetric spin--orbit coupling (ASOC) that leads to a splitting of the Fermi surface and, moreover, gives rise to the unique possibility of having admixtures of spin--singlet and spin--triplet pairing states. At present, however, the relative magnitude of both contributions to the superconducting order parameter is unknown.

In this letter, we propose that inelastic (Raman) light scattering provides a powerful tool to solve this problem and, in general, to investigate the underlying condensate in such parity--violated, non--centrosymmetric superconductors (NCS). This is because various choices of the photon polarization with respect to the location of the nodes on the Fermi surface allow one to draw conclusions about the node topology and hence the pairing
symmetry. An example for the success of such an analysis is the work by Devereaux {\it et al.} \cite{Devereaux:1994:01} in which the $d_{x^2-y^2}$--symmetry of the order parameter in cuprate superconductors could be directly traced back to the frequency--dependence of the electronic Raman spectra, measuring the pair--breaking effect. Therefore, our predictions of the polarization dependence of Raman spectra enable one to draw conclusions about the internal structure of the parity--mixed condensate in a given NCS.

\begin{figure} 
\begin{minipage}[c]{0.45\linewidth}
   \noindent \hspace{-1.0\linewidth} (a)\\
  \includegraphics[angle=0, width=0.9\linewidth]{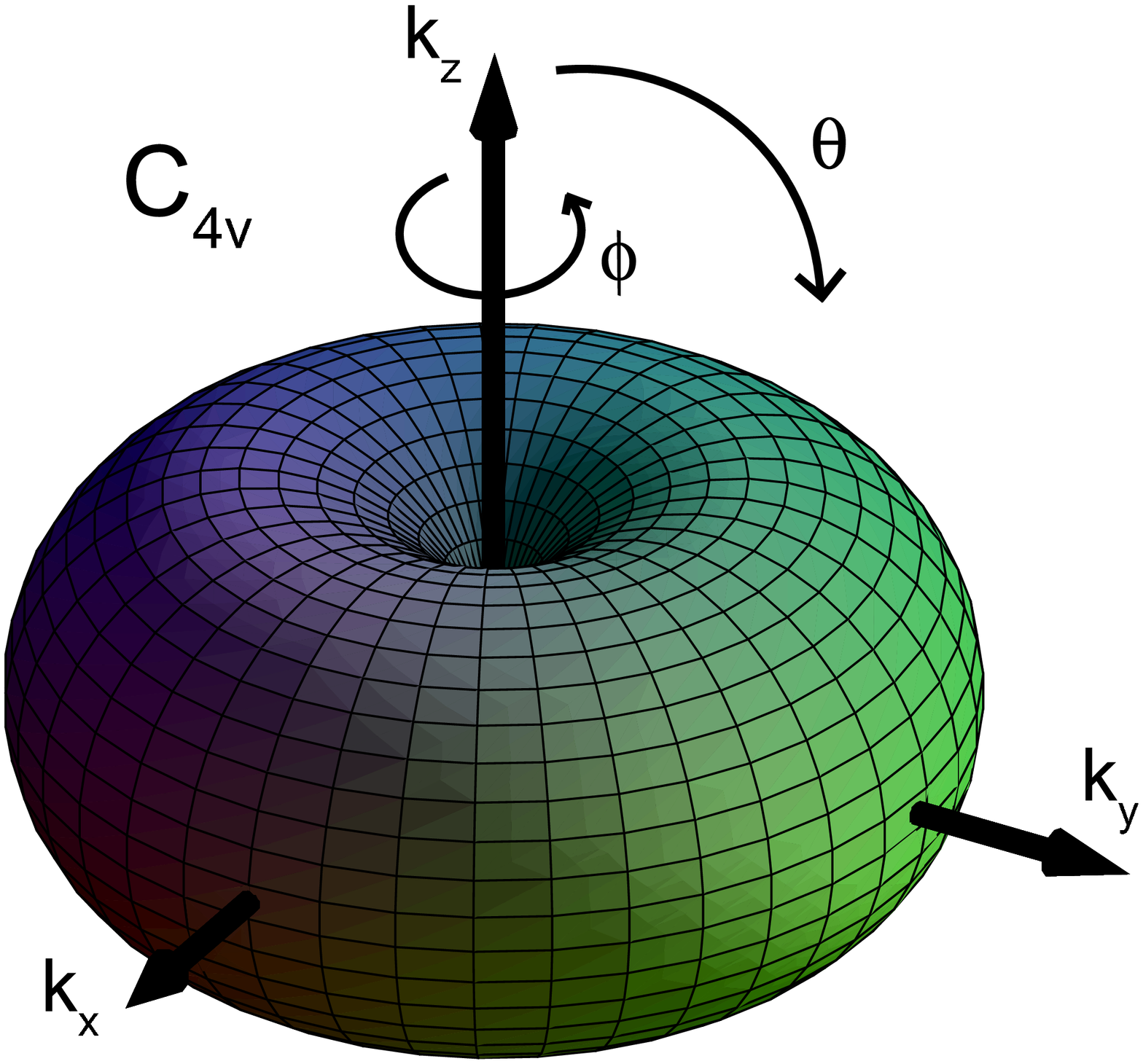}
\end{minipage}
\hspace{0.04\linewidth}
\begin{minipage}[c]{0.45\linewidth}
 \noindent \hspace{-1.0\linewidth} (b)\\
 \includegraphics[angle=0, width=0.9\linewidth]{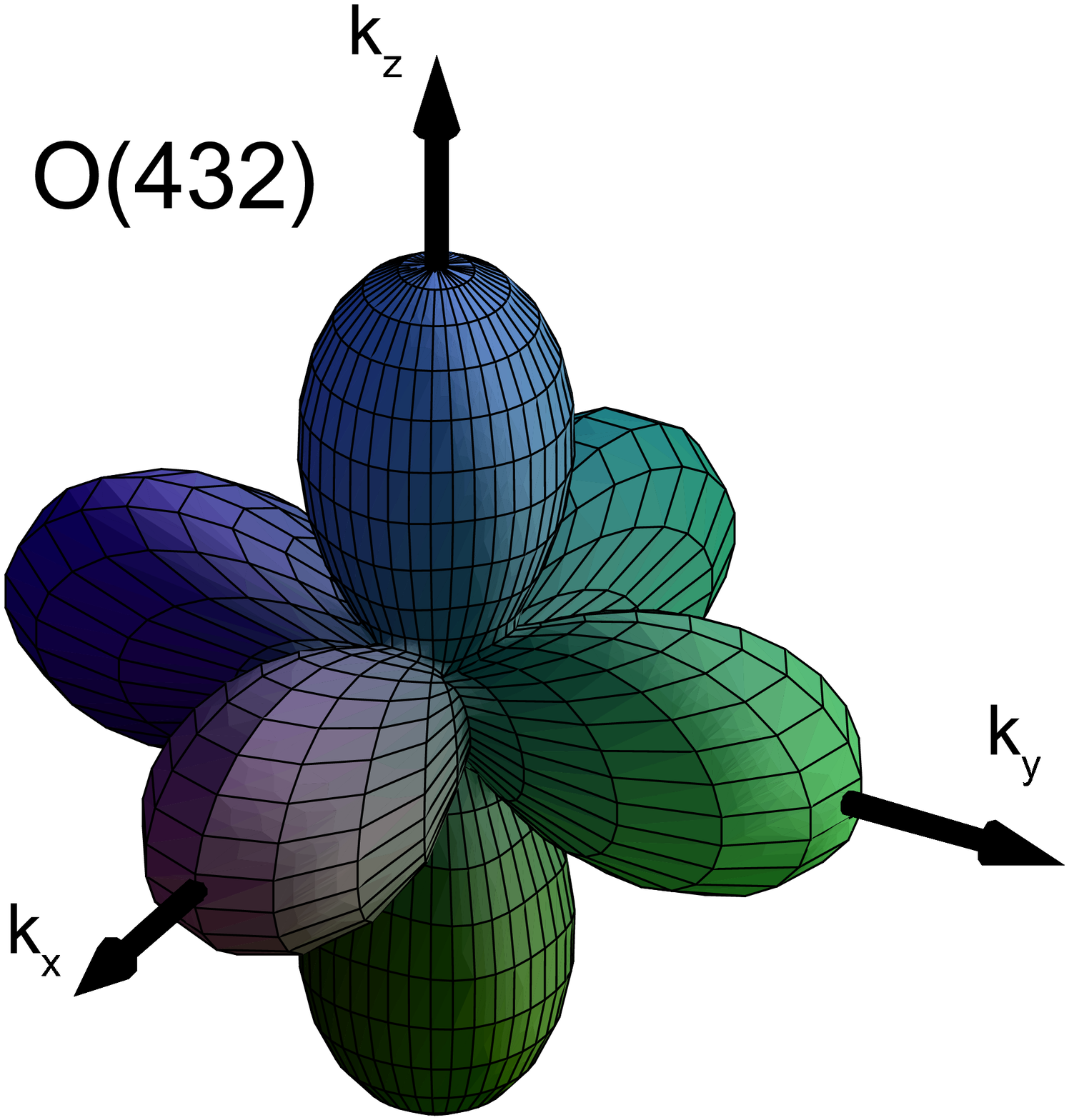}
\end{minipage}
\caption{(color online) The angular dependence of $|\vg_\vk|$ for the point groups C$_{4v}$ and O(423). Since $\vd_\vk || \vg_\vk$, these plots show also the magnitude of the gap function in the pure triplet case for both point groups.\label{Fig_gk}}
\end{figure}

\begin{figure} 
\begin{center}
\begin{minipage}[c]{0.05\linewidth}
  \rotatebox{90}{\fontfamily{phv}\selectfont \;\;\;\;\;\;Raman intensity (arbitrary units)}
\end{minipage}
\begin{minipage}[c]{0.575\linewidth}
 \includegraphics[angle=0, width=1.0\linewidth]{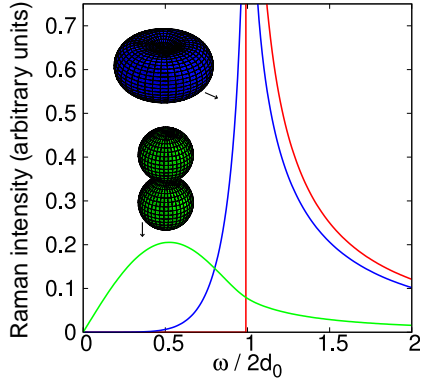}
\end{minipage}
\end{center}
\caption{(color online) Raman spectra for a pure triplet order parameter ($\psi=0$) for B$_{1,2}$ polarization of the point group C$_{4v}$ in backscattering geometry. The ABM (axial) state with $|\vd_\vk|=d_0 \sin \theta$ is displayed in blue and the polar state with $|\vd_\vk|=d_0 |\cos \theta|$ in green. For a comparison, also the Raman response for the BW state (red) with $|\vd_\vk|=d_0$ is shown.\label{Fig_triplet}}
\end{figure}

The model Hamiltonian for noninteracting electrons in a non-centrosymmetric crystal reads \cite{Samokhin:2007:01}
\begin{eqnarray}
\hat H &=& \sum_{\vk\sigma\sigma^\prime} \hat c^\dagger_{\vk\sigma}\left[ \xi_\vk \delta_{\sigma\sigma^\prime} + \vg_\vk\cdot\vtau_{\sigma\sigma^\prime} \right] \hat c_{\vk\sigma^\prime} \label{Eq_Hamilton}
\end{eqnarray}
where $\xi_\vk$ represents the bare band dispersion, $\sigma,\sigma^\prime=\uparrow,\downarrow$ label the spin state and $\vtau$ are the Pauli matrices. The second term describes an ASOC with a coupling $\vg_\vk$ \cite{footnote01}.
In NCS two important classes of ASOCs are realized reflecting the underlying point group $\mathcal{G}$ of the crystal. This is illustrated in Fig. \ref{Fig_gk}.
Therefore, we shall be interested in the tetragonal point group $C_{4v}$ (application to CePt$_3$Si, for example) and the cubic point group $O(432)$ (applicable to the system Li$_2$Pd$_x$Pt$_{3-x}$B).
For $\mathcal{G}=C_{4v}$ the ASOC reads \cite{Samokhin:2007:01}
\begin{align}
  \vg_\vk = \mathrm{g}_\bot ( \hat\vk\times\hat\ve_z ) + \mathrm{g}_\Vert \hat k_x\hat k_y\hat k_z ( \hat k^2_x - \hat k^2_y ) \hat\ve_z\;. \label{Eq_gk_tetragonal}
\end{align}
In the purely two-dimensional case ($\mathrm{g}_\Vert=0$) one recovers what is known as the Rashba interaction \cite{Edelstein:1989:01,Gorkov:2001:01}.
For the cubic point group $\mathcal{G}=O(432)$ $\vg_\vk$ reads \cite{Yuan:2006:01}
\begin{align}
  \vg_\vk &= \mathrm{g}_1 \hat \vk \label{Eq_gk_cubic} \\
  &- \mathrm{g}_3 \left[ \hat k_x (\hat k_y^2 + \hat k_z^2)\hat\ve_x + \hat k_y (\hat k_z^2 + \hat k_x^2)\hat\ve_y + \hat k_z (\hat k_x^2 + \hat k_y^2)\hat\ve_z \right] \nn
\end{align}
where the ratio $\mathrm{g}_3/\mathrm{g}_1\simeq3/2$ is estimated by Ref. \cite{Yuan:2006:01}.

Diagonalizing the Hamiltonian one finds the eigenvalues $\xi_{\vk\pm}=\xi_\vk\pm|\vg_\vk|$, which physically correspond to the lifting of the Kramers degeneracy between the two spin states at a given $\vk$.
Sigrist and co-workers have shown that the presence of the ASOC generally allows for an admixture between a spin--triplet order parameter and a spin--singlet pairing gap \cite{Frigeri:2004:02}. This implies that we can write down the following ansatz for the energy gap matrix in spin space
$\vDelta_{\vk\sigma\sigma^\prime} = [(\psi_\vk (T)\vone + \vd_\vk (T)\cdot \vtau) i \tau^y]_{\sigma\sigma^\prime}$, where $\psi_\vk(T)$ and $\vd_\vk(T)$ reflect the singlet and triplet part of the pair potential, respectively. It is then easy to see that the ASOC is not destructive for triplet pairing if one assumes $\vd_\vk \Vert \vg_\vk$ \cite{Frigeri:2004:02,Samokhin:2008:01}.
This results in the following ansatz for the gap function on both bands ($+$,$-$) \cite{Frigeri:2006:01}:
\begin{align}
  \Delta_{\vk\pm} = \psi \pm d|\vg_\vk| = \psi \left( 1 \pm p|\vg_\vk| \right) \equiv \Delta_\pm \label{Eq_gap}
\end{align}
where the parameter $p=d/\psi$ represents the unknown triplet--singlet ratio.
Note that for Li$_2$Pd$_x$Pt$_{3-x}$B this parameter seems to be directly related to the substitution of platinum by palladium, since the larger spin--orbit coupling of the heavier heavier platinum is expected to enhance the triplet contribution \cite{Lee:2005:01}. This seems to be confirmed by penetration depth experiments \cite{Yuan:2006:01}.

The T=0 electronic Raman response in a single band is given by the imaginary part of ($1\leftrightarrow\gamma_\vk=1$) \cite{Monien:1990:01}
\begin{eqnarray}
 \chi_{\gamma\gamma}(\omega) &=& \chi_{\gamma\gamma}^{(0)}(\omega) - \frac{\left[\chi^{(0)}_{\gamma 1}(\omega)\right]^2}{\chi^{(0)}_{11}(\omega)} \label{Eq_Raman_01}
\end{eqnarray}
where the index $\gamma=\gamma_\vk$ denotes the momentum--dependent Raman vertex that describes the coupling of polarized light to the sample. 
Note that the second term in Eq. (\ref{Eq_Raman_01}) is often referred to as the screening contribution that originates from gauge invariance. Since the ASOC leads to a splitting of the Fermi surface, the total Raman response is given by $\chi_{\gamma\gamma}^{\rm total} = \sum_{\lambda=\pm} \chi_{\gamma\gamma}^\lambda$ with $\chi^\pm_{\gamma\gamma}=\chi_{\gamma\gamma}(\Delta_\pm)$, in which the usual summation over the spin variable $\sigma$ is replaced by a summation over the pseudo--spin (band) index $\lambda$. With Eq. (\ref{Eq_gap}), the unscreened Raman response for both bands can be written as
\begin{eqnarray}
  \Im\chi_{\gamma\gamma}^{(0)\pm} = \frac{\pi\NF^\pm\psi}{\omega} \Re \left\langle \gamma_\vk^2 \frac{\left| 1\pm p|\vg_\vk| \right|^2}{\sqrt{(\frac{\omega}{2\psi})^2 - \left| 1\pm p|\vg_\vk| \right|^2}} \right\rangle_{\rm FS}
\end{eqnarray}
where we allow for a different density of states $\NF^\pm$ on both bands and $\langle\ldots\rangle_{\rm FS}$ denotes an average over the Fermi surface. 
For small momentum transfers and nonresonant scattering, the Raman tensor is given by $\vgamma_\vk = m \sum_{\alpha,\beta} \ve^S_\alpha (\partial^2 \epsilon(\vk)/\partial k_\alpha \partial k_\beta) \ve^I_\beta$
where $\ve^{S,I}$ denote the scattered and incident polarization light vectors.
The light polarization selects elements of this Raman tensor, where $\vgamma_\vk$ can be decomposed into its symmetry components and, after a lengthy calculation, expanded into a set of basis functions on a spherical Fermi surface.
Our results for the tetragonal group C$_{4v}$ are
\begin{align}
	\gamma_{A_1} &= \sum\limits_{k=0}^\infty \sum\limits_{l=0}^{l\leq \frac{k}{2}} \gamma_{k,l}\cos4l\phi\, \sin^{2k}\theta \label{Eq_vertex_C4v}\\
	\gamma_{B_1} &= \sum\limits_{k=1}^\infty \sum\limits_{l=1}^{l\leq \frac{k+1}{2}} \gamma_{k,l}\cos(4l-2)\phi\, \sin^{2k}\theta \nn\\
	\gamma_{B_2} &= \sum\limits_{k=1}^\infty \sum\limits_{l=1}^{l\leq \frac{k+1}{2}} \gamma_{k,l}\sin(4l-2)\phi\, \sin^{2k}\theta \nn
\end{align}
and for the cubic group $O(432)$ we obtain
\begin{align}
	\gamma_{A_1} &= \sum\limits_{k=0}^\infty \sum\limits_{l=0}^{l\leq \frac{k}{2}} \gamma_{k,l}\cos4l\phi\, \sin^{2k}\theta \label{Eq_vertex_O432}\\
	\gamma_{E^{(1)}} &= \gamma_0 (2-3\sin^2\theta) + \ldots \nn \\
	\gamma_{E^{(2)}} &= \sum\limits_{k=1}^\infty \sum\limits_{l=1}^{l\leq \frac{k+1}{2}} \gamma_{k,l}\cos(4l-2)\phi\, \sin^{2k}\theta \nn \\
	\gamma_{T_2} &= \sum\limits_{k=1}^\infty \sum\limits_{l=1}^{l\leq \frac{k+1}{2}} \gamma_{k,l}\sin(4l-2)\phi\, \sin^{2k}\theta \nn
\end{align}
in a backscattering--geometry experiment ($z\overline{z}$) \cite{footnote02}. In what follows, we neglect higher harmonics and thus use only the leading term in the expansions of $\vgamma_\vk$.

Before studying the mixed-parity case, it is instructive to analyze the Raman response for pure triplet $p$-wave pairing, see Fig. \ref{Fig_triplet}. Some representative examples are the Balian-Werthamer (BW) state, the Anderson-Brinkman-Morel (ABM or axial) state, and the polar state. The simple pseudoisotropic BW state with $\vd_\vk= d_0 \hat\vk$ [equivalent to Eq. (\ref{Eq_gk_cubic}) for $\mathrm{g}_3=0$], as well as previous work on triplet superconductors, restricted on a (cylindrical) 2D Fermi surface, generates the same Raman response as an $s$-wave superconductor \cite{Kee:2003:01}. In 3D we obtain more interesting results for the axial state with $\vd_\vk=d_0(\hat k_y\hat\ve_x - \hat k_x\hat\ve_y)$ [equivalent to Eq. (\ref{Eq_gk_tetragonal}) for $\mathrm{g}_\parallel=0$]. The Raman response for this axial state in B$_1$ and B$_2$ polarizations for $\mathcal{G}=C_{4v}$ is given then by
\begin{eqnarray}
 \chi^{\prime\prime}_{B_{1,2}} (x) &=& \frac{\pi \NF \gamma_0^2}{128} \left( -10 -\frac{28}{3}x^2-10x^4 \right. \label{Eq_triplet_ABM}\\
 &+& \left. \frac{5+3x^2+3x^4+5x^6}{x} \ln \left| \frac{x+1}{x-1} \right| \right) \nn
\end{eqnarray}
with the reduced frequency $x=\omega/2d_0$. An expansion for low frequencies reveals a characteristic exponent [$\chi^{\prime\prime}_{B_{1,2}} \propto \left(\omega/2d_0\right)^6$] which is due to the overlap between the gap and the vertex function. 
Moreover, we calculate the Raman response for the polar state with $\vd_\vk=d_0\hat k_z \hat\ve_x$, where one equatorial line node crosses the Fermi surface:
\begin{align}
 \chi^{\prime\prime}_{B_{1,2}} (x) = \frac{\pi \NF \gamma_0^2}{8x}
 \begin{cases}
    \frac{\pi}{2}x^2-\frac{3\pi}{4}x^4+\frac{5\pi}{16}x^6 & x\leq 1 \\
    \left( x^2-\frac{3}{2}x^4+\frac{5}{8}x^6 \right) \arcsin \frac{1}{x} & x > 1\\
      - \left( \frac{1}{3}-\frac{13}{12}x^2+\frac{5}{8}x^4 \right) \sqrt{x^2-1} 
 \end{cases} \nn
\end{align}
with the trivial low frequency expansion $\chi^{\prime\prime}_{B_{1,2}} \propto \omega/2d_0$.
Whereas the pair--breaking peaks for the BW and ABM state were both located at $\omega=2d_0$ (similar to the B$_{1g}$ polarization in the singlet $d$-wave case, which is peaked at $2\Delta_0$), for the polar state this peak is significantly shifted to lower frequencies. The maximum is given by $\omega=1.38 d_0$, which looks similar to the response for B$_{2g}$ polarization in singlet $d$-wave superconductors, where it is also peaked at $\omega<2\Delta_0$.

In general, due to the mixing of a singlet and a triplet component to the superconducting condensate, one expects a two--peak structure in parity--violated NCS, reflecting both pair--breaking peaks for the linear combination [see Eq. (\ref{Eq_gap})] of the singlet order parameter $\psi_\vk$ (extensively discussed in Ref. \cite{Devereaux:1995:01}) and the triplet order parameter $\vd_\vk$ (shown in Fig. \ref{Fig_triplet}), respectively. The ratio $p=d/\psi$, however, is unknown for both types of ASOCs.

\begin{figure} 
\begin{minipage}[c]{0.05\linewidth}
  \rotatebox{90}{\fontfamily{phv}\selectfont Raman intensity (arbitrary units)}
\end{minipage}
\begin{minipage}[c]{0.9\linewidth}
\setlength{\tabcolsep}{-0.09\linewidth}	
  \begin{tabular}{cc}
     \includegraphics[angle=-90, width=0.7\linewidth]{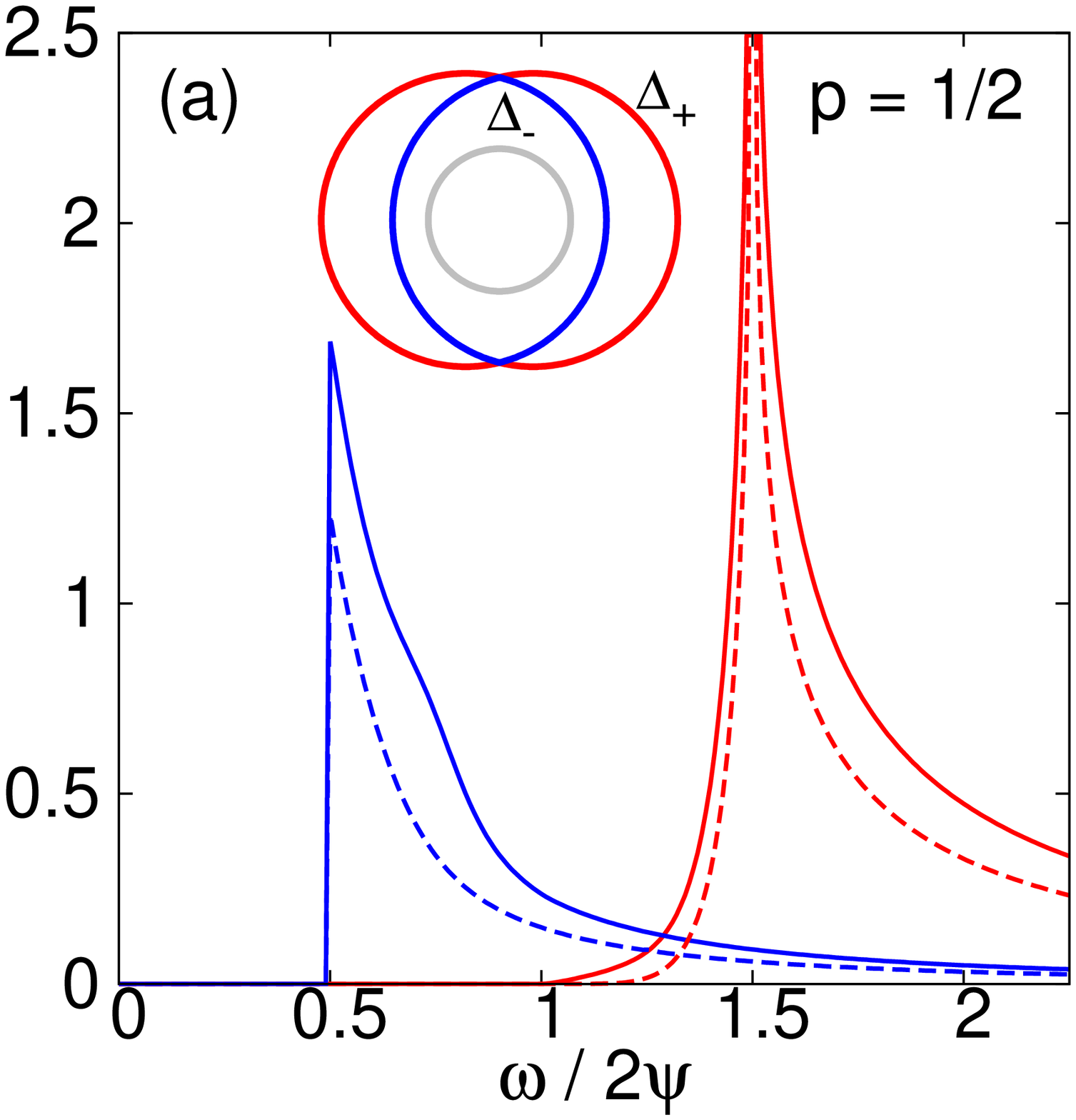} &
     \includegraphics[angle=-90, width=0.7\linewidth]{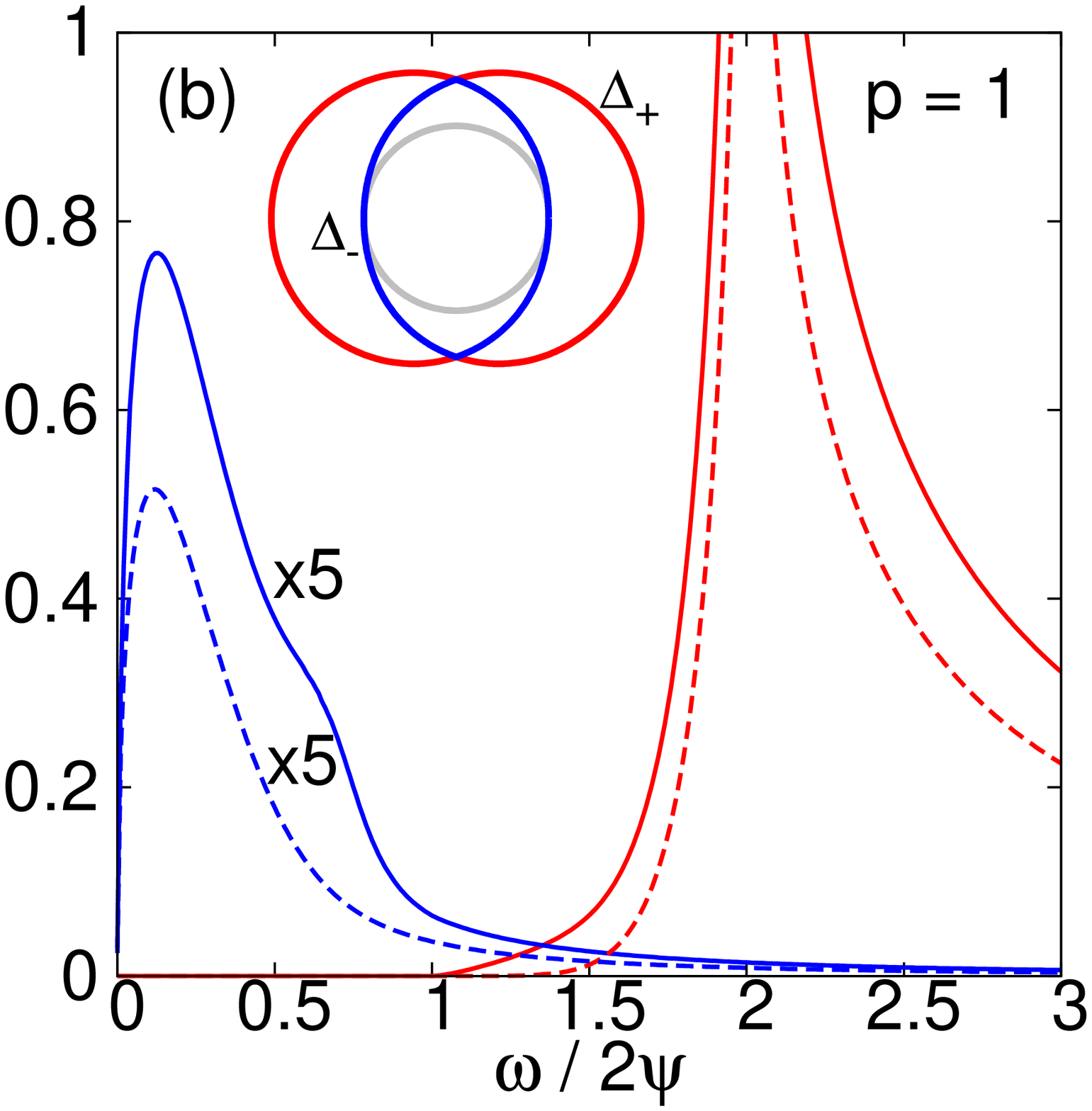} \\[18pt]
     \includegraphics[angle=-90, width=0.7\linewidth]{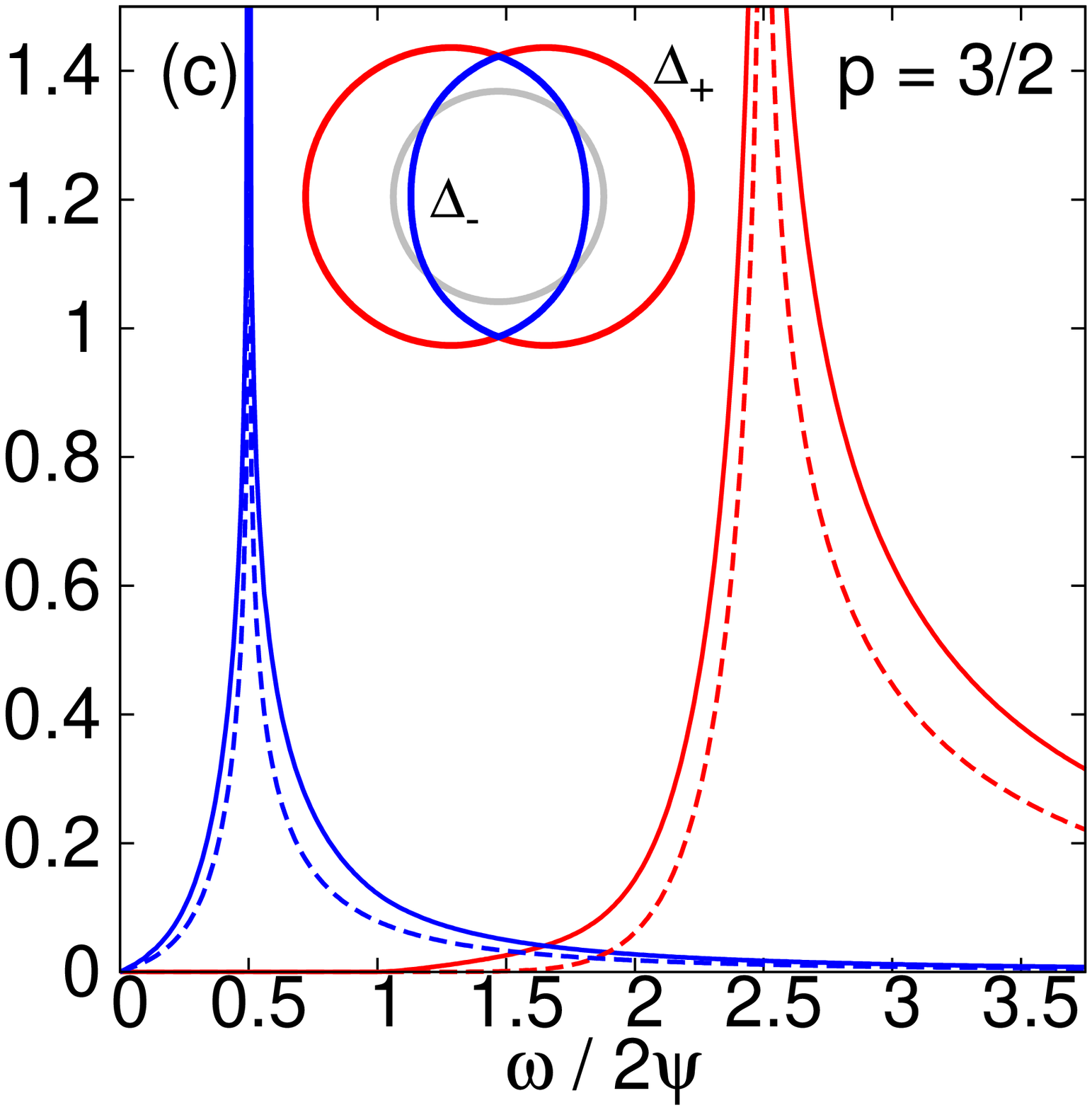} &
     \includegraphics[angle=-90, width=0.7\linewidth]{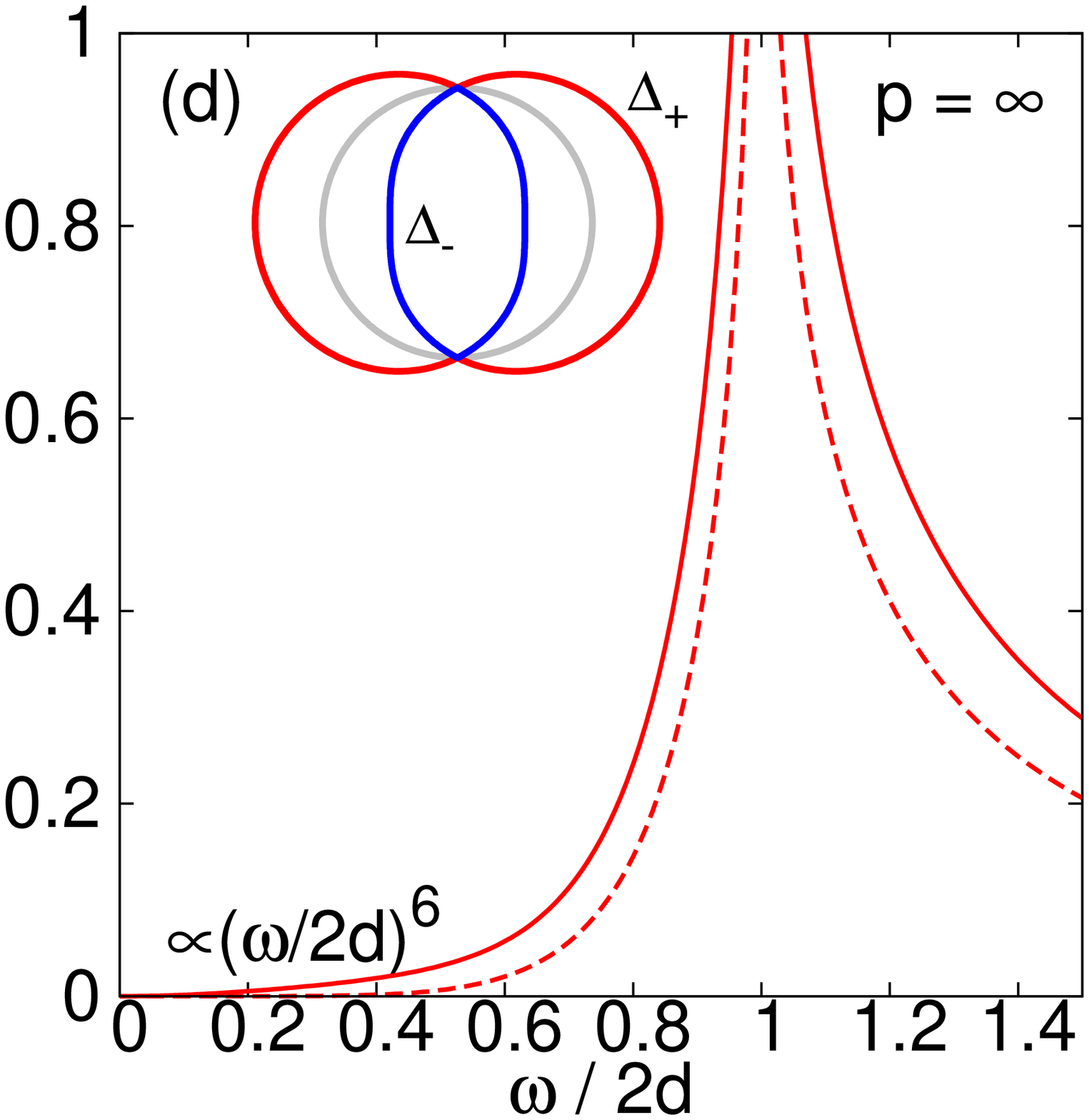} \\
  \end{tabular}
\setlength{\tabcolsep}{0mm}	
\end{minipage}
\caption{(color online) Theoretical prediction of the Raman spectra $\chi^{\prime\prime}_{\gamma\gamma}(\Delta_-)$ [blue] and $\chi^{\prime\prime}_{\gamma\gamma}(\Delta_+)$ [red] for A$_1$ (solid lines) and for B$_{1,2}$ (dashed lines) polarizations for the point group C$_{4v}$. We obtain the same spectra for the B$_1$ and B$_2$ symmetry. The polar diagrams in the insets demonstrate the four qualitative different cases.
\label{Fig_C4v}}
\end{figure}

In Fig. \ref{Fig_C4v} we show the calculated Raman response using Eq. (\ref{Eq_gk_tetragonal}) with $\mathrm{g}_\parallel=0$. This Rashba--type of ASOC splits the Fermi surface into two bands; while on the one band the gap function is $\Delta_{\vk} = \psi \left( 1 + p|\vg_\vk| \right)\equiv\Delta_+$, it is $\Delta_- \equiv \psi \left( 1 - p|\vg_\vk| \right)$ on the other band.
Thus, depending on the ratio $p=d/\psi$, four different cases (see polar diagrams in the insets) have to be considered: (a) no nodes; (b) one (equatorial) line node ($\Delta_-$ band); (c) two line nodes ($\Delta_-$ band); and (d) two point nodes on both bands.
Since the Raman intensity in NCS is proportional to the imaginary part of $\chi_{\gamma\gamma}^{\rm total}=\chi_{\gamma\gamma}(\Delta_-)+\chi_{\gamma\gamma}(\Delta_+)$, it is interesting to display both contributions separately (blue and red, respectively).
Even though (except for $\psi=0$) we always find two pair--breaking peaks at $\omega/2\psi=|1\pm p|$ we stress that our results for NCS are not just a superposition of a singlet and a triplet spectra. This is clearly demonstrated in Fig \ref{Fig_C4v}(a), for example, in which we show the results for a small triplet contribution ($p=1/2$). For $\chi_{\gamma\gamma}^{\prime\prime}(\Delta_-)$ we find a threshold behavior with an adjacent maximum value of $\chi^{\prime\prime}_{B_{1,2}}(\Delta_-)=\NF^-\gamma_0^2\,\pi^2/8\sqrt{p^{-1}-1}$ and for $\chi_{\gamma\gamma}^{\prime\prime}(\Delta_+)$ a zero Raman signal to twice the singlet contribution followed by a smooth increase and a singularity \cite{footnote03}.
For the special case (b), where the singlet contribution equals the triplet one ($p=1$), the gap function $\Delta_-$ displays an equatorial line node without sign change. Because of this nodal structure and strong weight from the vertex function ($\propto\sin^2\theta$), many low energy quasiparticles can be excited, which leads to this square--root--like increase in the Raman intensity. In this special case the pair--breaking peak is located very close to elastic scattering ($\omega=0.24\psi$).
In Fig. \ref{Fig_C4v}(c) the gap function $\Delta_-$ displays two circular line nodes. The corresponding Raman response for $p>1$ shows two singularities with different low frequency power laws [$\chi^{\prime\prime}_{B_{1,2}}(\Delta_-)\propto\omega/2\psi$ and $\chi^{\prime\prime}_{B_{1,2}}(\Delta_+)\propto(\omega/2\psi-1)^{\frac{11}{2}}$].
Finally, for $p\gg1$ one recovers the pure triplet cases (d) which is given analytically by Eq. (\ref{Eq_triplet_ABM}).

\begin{figure} 
\begin{minipage}[c]{0.05\linewidth}
  \rotatebox{90}{\fontfamily{phv}\selectfont Raman intensity (arbitrary units)}
\end{minipage}
\begin{minipage}[t]{0.9\linewidth}
\setlength{\tabcolsep}{-0.09\linewidth}	
  \begin{tabular}{cc}
     \includegraphics[angle=-90, width=0.7\linewidth]{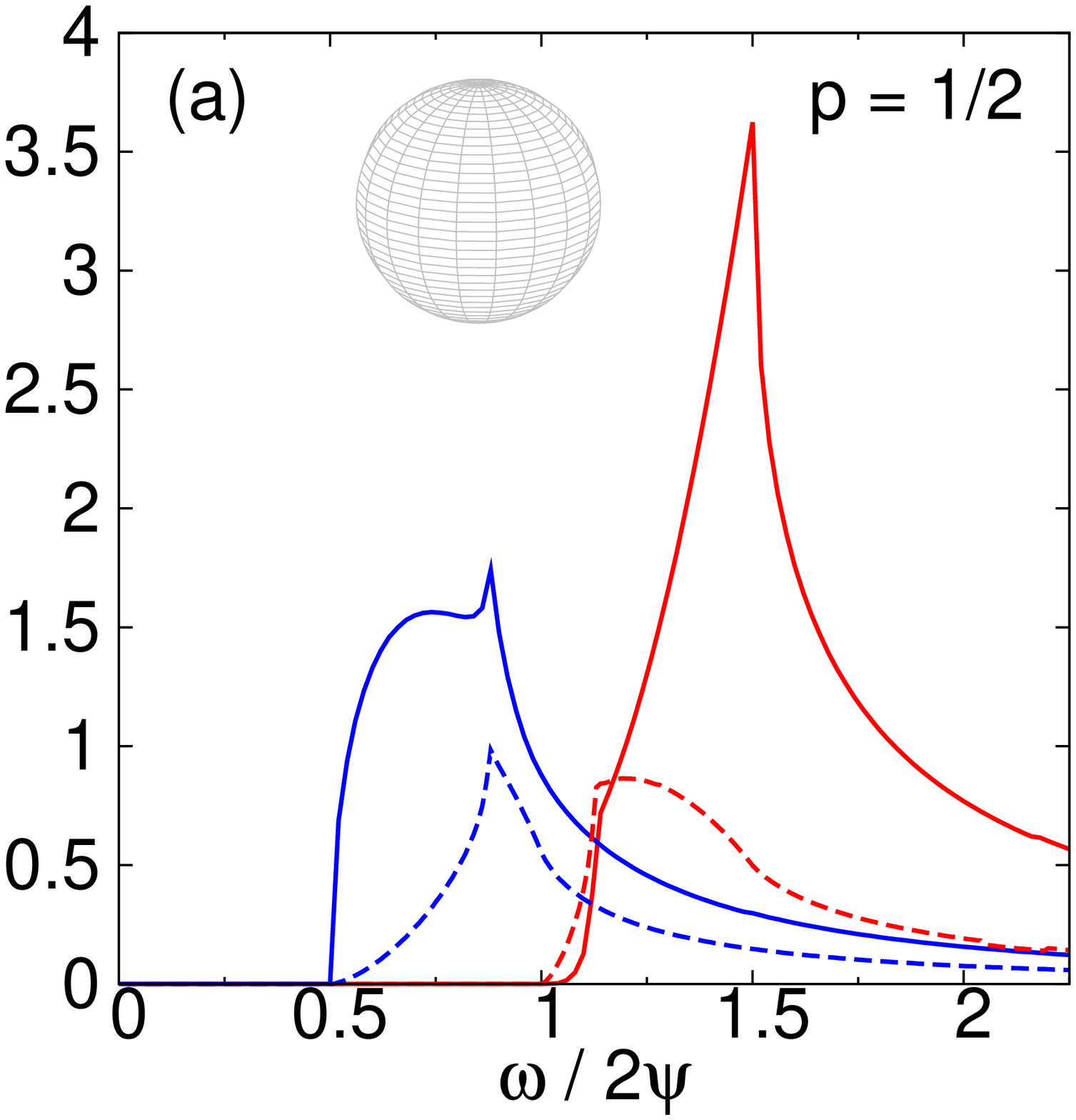} &
     \includegraphics[angle=-90, width=0.7\linewidth]{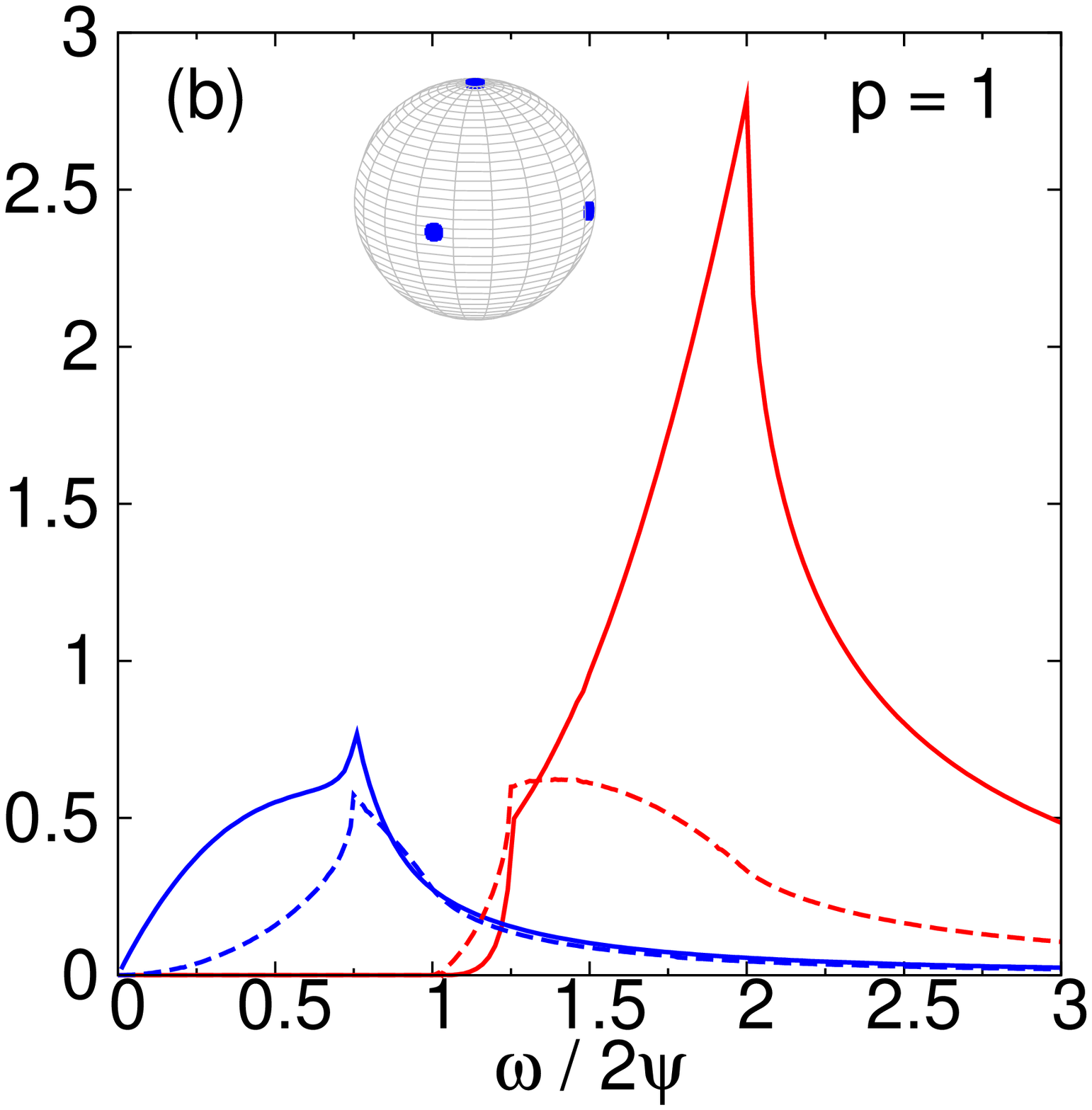} \\[18pt]
     \includegraphics[angle=-90, width=0.7\linewidth]{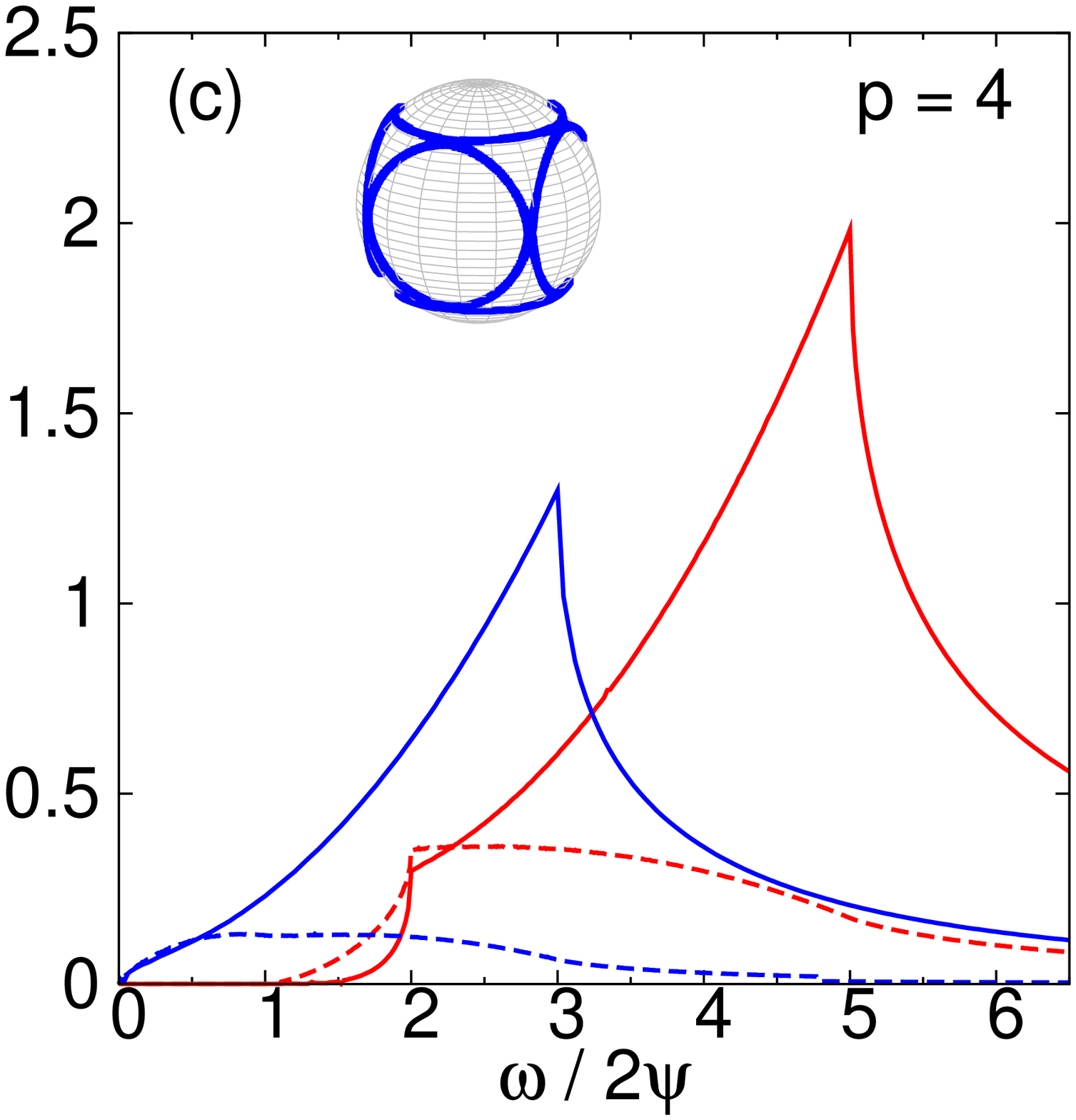} &
     \includegraphics[angle=-90, width=0.7\linewidth]{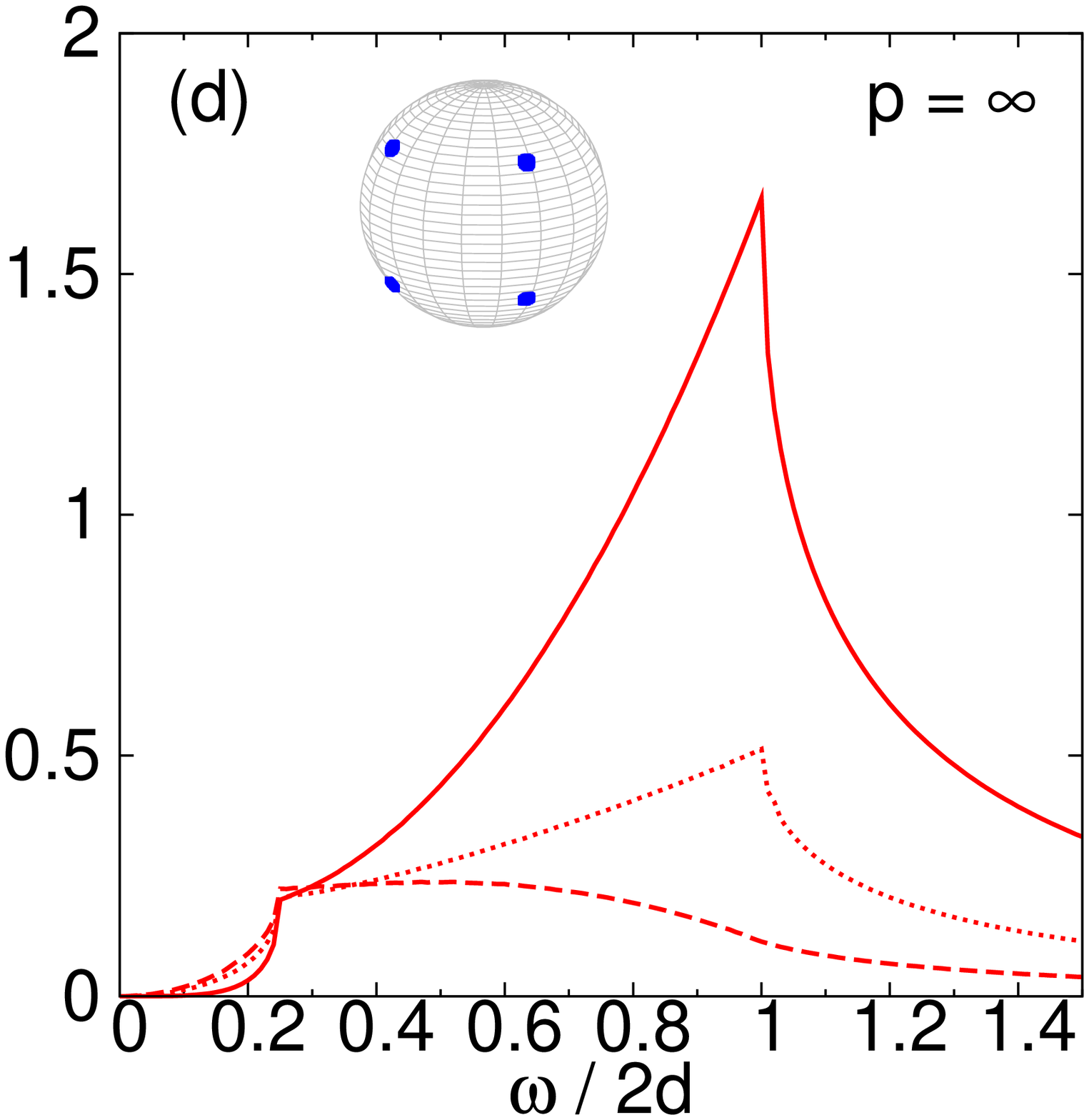} \\
  \end{tabular}
\setlength{\tabcolsep}{0mm}	
\end{minipage}
\caption{(color online) Theoretical prediction of the Raman spectra $\chi_{\gamma\gamma}(\Delta_-)$ [blue] and $\chi_{\gamma\gamma}(\Delta_+)$ [red] for E (solid lines), T$_2$ (dashed lines) and A$_1$ [dotted line, only in (d)] polarizations for the point group O(432). The insets display the point and line nodes of the gap function $\Delta_-$.\label{Fig_O432}}
\end{figure}

The Raman response for the point group $O(432)$, using Eq. (\ref{Eq_gk_cubic}), is shown in Fig. \ref{Fig_O432}.
We again consider four different cases: (a) no nodes; (b) six point nodes ($\Delta_-$ band); (c) six connected line nodes ($\Delta_-$ band); and (d) 8 point nodes (both bands) as illustrated in the insets.
Obviously, the pronounced angular dependence of $|\vg_\vk|$ leads to a strong polarization dependence. Thus we get different peak positions for the E and T$_2$ polarizations in $\chi_{\gamma\gamma}^{\prime\prime}(\Delta_+)$.
As a further consequence, the Raman spectra reveals up to two kinks on each band ($+$,$-$) at $\omega/2\psi=|1\pm p/4|$ and $\omega/2\psi=|1\pm p|$ \cite{footnote04}.
Furthermore, no singularities are present.
Nevertheless, the main feature, namely the two--peak structure, is still present and one can directly deduce the value of $p$ from the peak and kink positions.
Finally, for $p\gg1$ one recovers the pure triplet case (d), in which the unscreened Raman response is given by
\begin{eqnarray*}
  \chi^{\prime\prime}_{\gamma\gamma}(\omega) \propto \frac{2d}{\omega} \Re \left\langle \gamma^2_\vk \frac{|\vg_\vk|^2}{\sqrt{(\omega/2d+|\vg_\vk|)(\omega/2d-|\vg_\vk|)}} \right\rangle_{\rm FS}\;.
\end{eqnarray*}
Clearly, only the area on the Fermi surface with ${\omega/2d>|\vg_\vk|}$ contributes to the Raman intensity. Since $|\vg_\vk|\in [0,1]$ has a saddle point at $|\vg_\vk|=1/4$, we find kinks at characteristic frequencies $\omega/2d=1/4$ and $\omega/2d=1$.
In contrast to the Rashba--type ASOC, we find a characteristic low energy expansion $\propto(\omega/2d)^2$ for both the A$_1$ and E symmetry, while $\propto(\omega/2d)^4$ for the T$_2$ symmetry \cite{hugepaperKME}.

In summary, we have calculated for the first time the electronic (pair--breaking) Raman response in the newly discovered NCS such as CePt$_3$Si ($\mathcal{G}=C_{4v}$) and Li$_2$Pd$_x$Pt$_{3-x}$B ($\mathcal{G}=O(432)$). Taking the pronounced ASOC into account, we provide various analytical results for the Raman response function and cover all relevant cases from weak to strong triplet--singlet ratio $p$. Our theoretical predictions can be used to analyze the underlying condensate in parity--violated NCS and allow the determination of $p$.

\begin{acknowledgments}
We thank P.~M.~R. Brydon and M. Sigrist for helpful discussions.
\end{acknowledgments}


\begin{thebibliography}{18}
\bibitem{footnote01}
$\vg_\vk$ has the symmetry properties: $\vg_{-\vk}=-\vg_\vk$ and $g\vg(g^{-1}\vk)=\vg_\vk$, where $g$ is any symmetry operation of the point group $\mathcal{G}$ of the crystal.
\bibitem{footnote02}
The vertices E$^{(1)}$ and E$^{(2)}$ look completely different but the Raman response is exactly the same because E$^{(1)}$ and E$^{(2)}$ are both elements of the same symmetry class.
\bibitem{footnote03}
Note that even though the gap function does not depend on $\phi$ (see Fig. \ref{Fig_gk}), we obtain a small polarization dependence. This unusual behavior only in A$_1$ symmetry is due to screening and leads to a small shoulder for $p\leq 1$.
\bibitem{footnote04}
Interestingly, the T$_2$ symmetry displays only a change in slope at $\omega/2\psi=|1+p|$ instead of a kink. A detailed analysis can be found in \cite{hugepaperKME}.
\end{thebibliography}

\end{document}